\documentclass[apj,a4paper,12pt,useAMS]{emulateapj}

\usepackage{amsmath}
\usepackage{cases}
\usepackage{amssymb}
\usepackage{graphicx}
\usepackage [latin1]{inputenc}

\begin{document}

\title{Double Neutron Star Mergers: Are Late-time Radio Signals
Overestimated?}

\author{Shao-Ze Li$^{1,5,6}$, Yun-Wei Yu$^{2}$, He Gao$^{3,4}$, and Lin
Lan$^{3,4}$}

\altaffiltext{1}{College of Physics Science and Technology, Hebei
University, Baoding 071002, China; lishaoze@mails.ccnu.edu.cn}
\altaffiltext{2}{Institute of Astrophysics, Central China Normal
University, Wuhan 430079, China; yuyw@mail.ccnu.edu.cn}
\altaffiltext{3}{Institute for Frontier in Astronomy and
Astrophysics, Beijing Normal University, Beijing 102206, China;
gaohe@bnu.edu.cn} \altaffiltext{4}{Department of Astronomy, Beijing
Normal University, Beijing 100875, China} \altaffiltext{5}{Hebei Key
Laboratory of High-precision Computation and Application of Quantum
Field Theory, Baoding, 071002, China} \altaffiltext{6}{Hebei
Research Center of the Basic Discipline for Computational Physics,
Baoding, 071002, China}

\begin{abstract}

The coalescence of binary neutron stars can yield the expulsion of a
fast-moving, quasi-isotropic material, which may induce thermal
radiation and give rise to kilonova emission. Moreover, the
interaction between the ejected material and the surrounding
environment generates an external shock, which can result in a
long-lasting radio signal that persists for several decades
following the merger. In contrast to supernova ejecta, the kilonova
ejecta exhibits a relatively lesser mass and higher velocity, and
its expansion may ultimately result in the ejecta density becoming
so low that the medium particles can freely pass through the ejecta.
It would thereby lead to a kind of incomplete sweeping on the
interstellar medium. Employing a toy model, our investigation
reveals that such incomplete sweeping may considerably diminish the
late-time radio radiation power, irrespective of whether the binary
neutron star merger results in the formation of a black hole or a
neutron star. Our findings thus imply that the previously reported
radio upper limits for certain short gamma-ray bursts (GRBs) may not
necessarily place stringent constraints on the presence of a
long-lived magnetar remnant in these short GRBs.

\end{abstract}

\section{Introduction}

The first detection of a gravitational-wave (GW) signal from a
merger of double neutron stars (NSs), GW170817, opens new era of
mutimessenger astronomy (Abbott et al. 2017a). The association of
the GW, the short gamma-ray burst (GRB), and the kilonova signals
confirms that at least some of the short GRBs come from mergers of
NSs (Abbott et al. 2017b; Cowperthwaite et al. 2017; Nicholl et al.
2017; Pian et al. 2017; Smartt et al. 2017; Tanaka et al. 2017;
Tanvir et al. 2017; Troja et al. 2017; Villar et al. 2017). It has
been almost 6 yr since the detection of GW170817. However, it is
still puzzling whether the merger remnant is a black hole (BH) or a
massive NS (Sarin \& Lasky 2021). In the literature, a millisecond
magnetar is usually proposed as one candidate of the central engine
of GRBs (Dai \& Lu 1998a, 1998b; Zhang \& M\'{e}sz\'{a}ros 2001;
Metzger et al. 2011). Compared with the BH model, the magnetar model
is helpful in understanding the observational features, such as the
plateau and flares in the GRB afterglow emission (Vaughan et al.
2006; Falcone et al. 2007; Rowlinson et al. 2013; L\"{u} et al.
2015; Gao et al. 2016). Meanwhile, the energy injection from the
central magnetar powers the isotropic merger ejecta, resulting in
the much brighter kilonova emission (mergernova; Yu et al. 2013,
2015, 2018; Metzger \& Piro 2014; Gao et al. 2015, 2017; Metzger et
al. 2017; Li et al. 2018; Ai et al. 2022; Sarin et al. 2022).

If the post-merger remnant is indeed a long-lived massive NS, rather
than a BH, it would give stringent constraints on the equation of
state of dense matter (Gao et al. 2016; Margalit \& Metzger 2017;
Shibata et al. 2017, 2019; Ai et al. 2018; Radice et al. 2018;
Rezzolla et al. 2018; Ai et al. 2020). Metzger \& Bower (2014)
proposed that the isotropic radio afterglow of the merger ejecta can
be used to test the merger remnant. Following the explosion of the
ejecta, an external shock is formed and sweeps up the surrounding
medium. Depending on specific parameters, the external shock could
give a long-lasting radio signal several years or several tens of
years after the merger (Nakar \& Piran 2011). Due to the existence
of the magnetar, the merger ejecta can be accelerated into
relatively higher velocity, which results in much brighter radio
emission (Gao et al. 2013). Based on this starting point, several
works have tried to catch the radio signals from the GW170817 event
and nearby short GRBs, but no signals were detected (Metzger \&
Bower 2014; Fong et al. 2016; Horesh et al. 2016; Klose et al. 2019;
Schroeder et al. 2020; Bruni et al. 2021; Ricci et al. 2021). Their
results give constraints on the central object, which seems to rule
out a magnetar remnant with rotation energy beyond $10^{52}\rm
ergs$. Nevertheless, Liu et al. (2020) point out that the magnetic
field fraction factor $\epsilon_{\rm B}$ is significant in
calculating the radiation mechanism. Taking a relatively lower
$\epsilon_{\rm B}$ derived from GRB afterglows, they obtain the
radio light curves using a more detailed model. The result shows
that a magnetar remnant with rotation energy $\sim 10^{52}\rm ergs$
is still allowed in broad parameter space.

In comparison to the supernova ejecta, the kilonova ejecta exhibits
a significantly lesser mass and higher velocity. Moreover, Margalit
\& Piran (2020) proposed a relativistic jetted outflow in front of
the ejecta, which would result in much earlier radio signals. No
matter the ejecta or the jetted outflow, their density could
eventually reach a low level as they expand to a certain extent.
This would make the ejecta or outflow so rarefied that it would have
hardly any significant interaction with the interstellar medium
(ISM). Meanwhile, it may result in a kind of incomplete sweeping on
the ISM. This effect has not previously been examined. To address
the incomplete sweeping, we first discuss the free path of the ISM
particles (ionized or neutral hydrogens) in Section 2. Subsequently,
we demonstrate the limitations of the complete-sweeping assumption
under typical conditions and construct an adapted toy model
incorporating the incomplete-sweeping effect (Section 3). In Section
4, we reevaluate the dynamics of the ejecta and illustrate the
discernible distinctions in the light curves of the late-time radio
emission. Our concluding remarks and discussions are presented in
Section 5.

\section{the free path}

In many astrophysical explosion events, such as supernovae, GRBs,
etc., the shock process usually plays a significant role in their
dynamics as well as the emissions. As extensively discussed in the
literature, a significant fraction of astrophysical shocks are
collisionless, especially in low-density flows. Such shocks are
characterized by plasma interactions instead of particle collisions.
For a charged particle moving within a plasma, the free path is
affected by the collective field, which operates on the scale of the
plasma skin depth,
\begin{eqnarray}
l_{\rm pl} &=& {c\over \omega_{\rm pe}}\\
&=&5.3\times  10^{5}\, n_{\rm e}^{-1/2} \rm cm,
\end{eqnarray}
where $\omega_{\rm pe}$ is the electron plasma frequency. Such a
free path is usually much less than the astrophysical flow scale. As
a result, collisionless shocks can form during the interactions of
plasma flows only if the relative velocity exceeds the local sound
speed. But not all astrophysical flows are plasmas. If the flows
consist of neutral molecules or atoms, there will be another
scenario. For a neutral particle, the free path is dominated by
elastic collisions:
\begin{eqnarray}
l_{\rm co} &=& (\sigma n_{\rm ej})^{-1}\nonumber\\
 &\approx& 10^{15} n_{\rm e}^{-1} \rm cm,\label{lco}
\end{eqnarray}
where $\sigma\approx10^{-15}\rm cm^2$ is approximately the cross
section of typical air molecules on the earth. Obviously, this free
path is much longer than that in plasma. The precondition for the
formation of a steady shock is $l\ll L$, which requires that the
scale of flows must be large enough. A typical example of this
scenario is the collisions between neutral molecular clouds.

In supernova-like events, it is usually assumed that the isotropic
ejecta would interact with the ambient ISM to form an external shock
in the ISM and sweep up all the surrounding medium. Such a
complete-sweeping assumption is commonly adopted in many studies to
estimate the long-lasting radio emission from kilonova events
(Metzger \& Bower 2014; Fong et al. 2016; Horesh et al. 2016; Liu et
al. 2020; Bruni et al. 2021; Ricci et al. 2021). However, despite
their similar dynamics, kilonovae and supernovae are different in
many aspects. The kilonova explosion has a much lesser ejecta mass
$\sim 0.01 M_{\odot}$, but with a relatively larger velocity $\sim
0.3 c$. This results in a particular diffusion time, $t_{\rm
d}\propto (M/v)^{1/2}$, which is roughly one day, corresponding to
the peak time of kilonova emission. Generally speaking, such a
diffusion timescale implies that the kilonova emission should be
fast-evolving, and the duration should be much shorter than that of
supernovae. Besides, supernovae and kilonovae also show differences
in the ambient ISM. In case of the progenitor star, core-collapse
supernovae usually locate at the H II regions. The continuous UV
emission from a bright star can ionize the hydrogen atoms and form
the so-called H II bubble. The typical scale for the H II bubble of
an OB star is $70\,\rm pc$. Such a kind of ISM is mostly composed of
ionized hydrogens. However, for the mergers of double NSs, they are
probably far away from the star-forming region. The ISM densities
derived from short-GRB afterglows range from $10^{-4}-1\rm cm^{-3}$
(Metzger \& Bower 2014), which are significantly lower than those in
H II region. Recently, Ag{\"u}{\'\i} Fern{\'a}ndez et al. (2023)
reported the first chemical study of the ISM in GRB160410A. On the
one hand, the X-ray spectrum shows strong Ly$\alpha$ absorption
lines, and the column density of $\rm log(N(HI)/cm^{2})\simeq 21.2$
indicates a damped Ly$\alpha$ host. This is similar to the long GRBs
having $\rm log(N(HI)/cm^{2})> 20.3$ (Tanvir et al. 2019). On the
other hand, the spectrum does not show any ionized $\rm C_{\rm \,
IV}$ and $\rm Si_{\rm \, IV}$ absorption lines. These lines detected
in long-GRB afterglows have been assumed to originate from the
ionized hot gas in the host galaxy (Ledoux et al. 1998; Wolfe \&
Prochaska 2000; Heintz et al. 2018). The result implies that this
short GRB is still within the galactic H I halo. Therefore, due to
the absence of the H II region and the hot diffuse gas, the merger
events are likely to be exposed to the neutral H I medium.

If the double NS merger events are indeed within the H I halo, the
fast-moving merger ejecta would interact with the ambient ISM, which
is mostly composed of neutral hydrogens. In this picture, the free
path for the neutral hydrogen atoms should be first dominated by
elastic collision.\footnote{Only if the neutral hydrogen atoms are
ionized during the shock can the free path be dominated by the
plasma effect.} Years after the explosion, the merger ejecta could
become so rarefied that the free path for neutral hydrogen atoms may
exceed the scale of the ejecta. As a result, a steady shock is
unsustainable, and most of the neutral ISM hydrogens would be
neither collided nor swept up by the rarefied ejecta. In this view,
the complete-sweeping assumption may be suspicious, especially at
late time.

After the nucleosynthesis during the merger explosion, the ejecta is
mostly composed of r-process elements, which are heavy metal
elements with mass numbers 100-200 (Rosswog et al. 2014; Hotokezaka
et al. 2018). After months and years of radioactive decay, these
elements mostly convert into iron-peak elements. Based on $^{56}{\rm
Fe}$, a mass number of $A_{\rm r}=150$ is adopted to characterize
these elements, for simplicity. The standard value is $m=1.66\times
10^{-24}\rm g$ for one-twelfth of the mass of $^{12}\rm C$. Then the
ejecta number density for supernovae and kilonovae can be estimated
by
\begin{eqnarray}
n_{\rm sn}&=& M_{\rm ej}/m V_{\rm ej}\nonumber\\
&=&9.2 \times 10^{4}\left({M_{\rm ej}\over
10M_{\odot}}\right)^{}\left({v_{\rm ej}\over 10^{4}{\rm
km\,s^{-1}}}\right)^{-3}\left({t_{\rm }\over 10 \rm
year}\right)^{-3}\rm cm^{-3},\label{nesn}
\end{eqnarray}
and
\begin{eqnarray}
n_{\rm kn}&=& M_{\rm ej}/A_{\rm r}m V_{\rm ej}\nonumber\\
&=&8.4 \times 10^{-4}\left({M_{\rm ej}\over
0.01M_{\odot}}\right)^{}\left({v_{\rm ej}\over
0.3c}\right)^{-3}\left({t_{\rm }\over 10 \rm year}\right)^{-3}\rm
cm^{-3}\label{nekn}
\end{eqnarray}
where $M_{\rm ej}$ is the mass, $V_{\rm ej}=4\pi R_{\rm ej}^{3}/3$
is the volume, $R_{\rm ej}= v_{\rm ej} t$ is the radius and $v_{\rm
ej}$ is the velocity, respectively. For the density of the supernova
ejecta, it is assumed that the ejecta is composed of hydrogens from
the stellar envelope of the progenitor. Obviously, the number
density of the merger ejecta years later is sufficiently lower than
that of the supernova ejecta and even lower than the typical ISM
density $\sim 1\rm cm^{-3}$. Actually, the merger ejecta is more
likely a kind of diffuse cloud that has already left the nebula
phase. Combining Equations (3) and (5), one can get the free path
for neutral hydrogen atoms:
\begin{eqnarray}
l &=& 1.2 \times 10^{19}\nonumber\\
 & &\left({\sigma \over  10^{-16}\rm
cm^2}\right)^{-1}\left({M_{\rm ej}\over
 0.01M_{\odot}}\right)^{-1}\left({v_{\rm ej}\over
 0.3c}\right)^{3}\left({t_{\rm }\over 10 \rm year}\right)^{3} \rm cm
 ,\label{freepath}
\end{eqnarray}
where the cross section for ejecta elements is adopted from solid
iron ($^{56}{\rm Fe}$). Assuming an average density of
$\overline{\rho}_{\rm }=10\rm g\, cm^{-3}$, the volume for each atom
is $\overline{V}_{\rm }= A_{\rm r}m / \overline{\rho} $. Considering
the close packing of metallic atoms, the cross section is then
roughly\footnote{It is noteworthy that the atomic cross section
utilized in our calculation is derived from the $^{56}{\rm Fe}$ atom
under standard conditions, which may introduce uncertainty in the
astrophysical setting.} $\sigma = \overline{V}_{\rm }^{2/3}\nonumber
= 8.5 \times 10^{-16}\rm cm^{2}. \label{sigma} $ This is slightly
smaller than the cross section of air molecules. From Equation (6),
we can see that the condition $l\ll L$ could be well satisfied at
early time with a relatively higher ejecta density. At late time,
one can see that the free path $l \propto t^{3}$ and the radius $R
\propto t^{}$. This suggests that the free path will finally exceed
the ejecta scale. As a natural consequence, there will be no steady
shock at late time, and the ejecta can hardly sweep up the ISM.
Therefore, when calculating the shock emission, it is significant to
ensure that the ejecta has the ability to sweep up the surrounding
medium.

\section{the sweeping}

\begin{figure}[t]
\centering\resizebox{1.0\hsize}{!}{\includegraphics{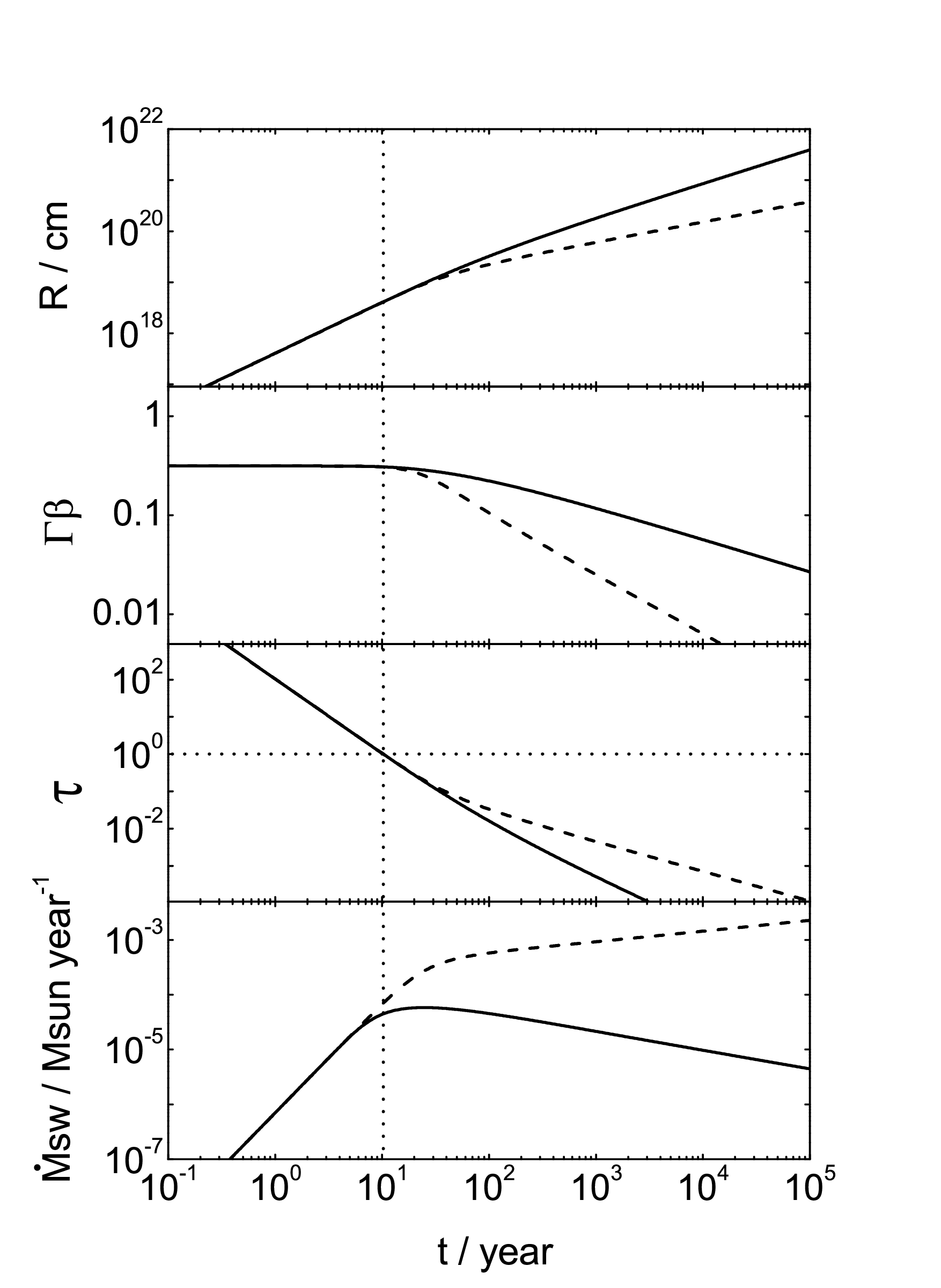}}\caption{The
dynamic evolution of the complete-sweeping (dashed lines) and
incomplete-sweeping (solid lines) scenarios. The dotted lines show
the location where $\tau=1$. The basic initial parameters are fixed
as $M_{\rm ej}= 0.01M_{\odot}$,
 $\beta_{\rm }=0.3$, $n=10^{-3}$, $p=2.2$, $\epsilon_{\rm e}=0.1$ and $\epsilon_{\rm B}=0.1$.}\label{fig1}
\end{figure}

Regardless of whether the shock could form, the interaction between
the merger ejecta and the ISM first requires that the free path
$l\ll L$. Therefore, it is of great significance to determine
whether the ambient ISM is ionized or not. Due to the natal kick of
the newborn NSs, the binary systems and indeed the merger events
more probably locate at the outskirts of the host galaxy, away from
the dense H II region. Conservatively, according to Ag{\"u}{\'\i}
Fern{\'a}ndez et al. (2023), it can be inferred that at least a
fraction of short GRBs are still located within the H I halo of
their host galaxies. In the subsequent calculations, it is assumed
that the merger events are directly within the H I halo, and the
ambient ISM is composed of pure neutral hydrogens.\footnote{The
jetted outflow proposed in Margalit \& Piran (2020) is not taken
into account in this analysis. If the outflow does exist, it can be
regarded as an ionized medium.} Basically, when the ejecta interacts
with the ISM, the neutral ISM hydrogens will be first collided by
the ejecta, ionized, and then shocked.\footnote{It is still a
collisionless shock in this case, but having first undergone a
collisional ionization.} Otherwise, the ejecta would continuously
move and diffuse in the neutral ISM and no steady shock could form.

In the beginning, the explosion starts at a small volume. The ejecta
density is high enough to sweep up all the ambient ISM. Then, the
deceleration timescale is
\begin{eqnarray}
t_{\rm dec} &=& {({3M_{\rm ej}/ 4\pi n_{\rm }})^{1/3}\over v_{\rm
ej}}\nonumber\\
&=& 8.7 \times \left({n_{\rm }\over 0.01\,\rm
cm^{-3}}\right)^{-1/3}\left({M_{\rm ej}\over
0.01M_{\odot}}\right)^{1/3}\left({v_{\rm ej}\over 0.3c}\right)^{-1}
{\rm year},
\end{eqnarray}
where $n$ is the number density of the surrounding medium. To
describe the sweeping, one can define a dimensionless parameter with
the free path of Equation (6):
\begin{eqnarray}
\tau &=& {R\over l}\nonumber\\
 &=& 0.24  \left({\sigma \over  10^{-16}\rm
cm^2}\right) \left({M_{\rm ej}\over
0.01M_{\odot}}\right)^{}\left({v_{\rm ej}\over
0.3c}\right)^{-2}\left({t_{\rm }\over 10 \rm year}\right)^{-2}
.\label{zeta}
\end{eqnarray}
When $\tau > 1$, the ejecta can be called thick; otherwise, the
ejecta is thin. The ejecta will finally become thin following the
explosion. One can define the critical time for such a transition,
which is
\begin{eqnarray}
t_{\tau}&=& 4.8\left({\sigma \over 10^{-16}\rm
cm^2}\right)^{1/2}\left({M_{\rm ej}\over
0.01M_{\odot}}\right)^{1/2}\left({v_{\rm ej}\over
0.3c}\right)^{-1}\rm year\label{ttau}.
\end{eqnarray}
We can see that the ejecta will become thin several years after the
merger. It is sensitive to the parameters when the transition
happens. By equating these two timescales, one can get the critical
density:
\begin{eqnarray}
n_{\rm c}&=& 0.057\left({M_{\rm ej}\over
0.01M_{\odot}}\right)^{-1/2}\left({\sigma \over 10^{-16}\rm
cm^2}\right)^{-3/2}.\label{nc}
\end{eqnarray}
When the density of the ambient ISM has $n_{\rm }>n_{\rm c}$, the
ejecta would still be thick at the deceleration timescale. That is
to say the ejecta would sweep up most of the surrounding medium and
be sufficiently decelerated. Otherwise, when $n_{\rm }<n_{\rm c}$,
the ejecta would become thin before the deceleration timescale,
which leads to the ejecta being unlikely to sweep up all the medium
and keeping its velocity for a much longer timescale.

The previous complete-sweeping assumption is invalid once the ejecta
becomes thin, i.e., the free path $l > R$. This means that the
expectation of the collision probability for neutral hydrogen atoms
is less than 1 during the interaction between the ejecta and the
ISM. To estimate the mass of the swept-up medium in the
incomplete-sweeping scenario, we use a modified equation,
\begin{eqnarray}
{dM_{\rm sw}\over dR} = 4\pi R^2  n_{\rm } m_{\rm p}(1-e^{-\tau})
,\label{dMswtau}
\end{eqnarray}
where $M_{\rm sw}$ is the total mass of the swept-up medium. The
advantage of this equation is that the surrounding medium can be
completely swept up when $\tau \gg 1$, but hardly swept up when
$\tau \ll 1$. Then, ignoring the energy lost by radiation, the
dynamics can be expressed by
\begin{eqnarray}
{d\Gamma\over dR} = -{(\Gamma^2-1 ) 4\pi R^2 n_{\rm } m_{\rm
p}(1-e^{-\tau})  \over M_{\rm ej} + 2 \Gamma M_{\rm sw}
},\label{dGamma}
\end{eqnarray}
together with
\begin{eqnarray}
{dR_{\rm }\over dt_{\rm }} = {\beta c \over 1-\beta},\label{dR}
\end{eqnarray}
where $\Gamma$ is the Lorentz factor. Note that, the dimensionless
parameter $\tau^{}=(R^{}/\Gamma)/(l^{}/\Gamma)=R/l$ is unchanged
during the Lorentz transformation. With above equations, the dynamic
equations can be solved out once the initial parameters are given.

\begin{figure}[t]
\centering\resizebox{1.0\hsize}{!}{\includegraphics{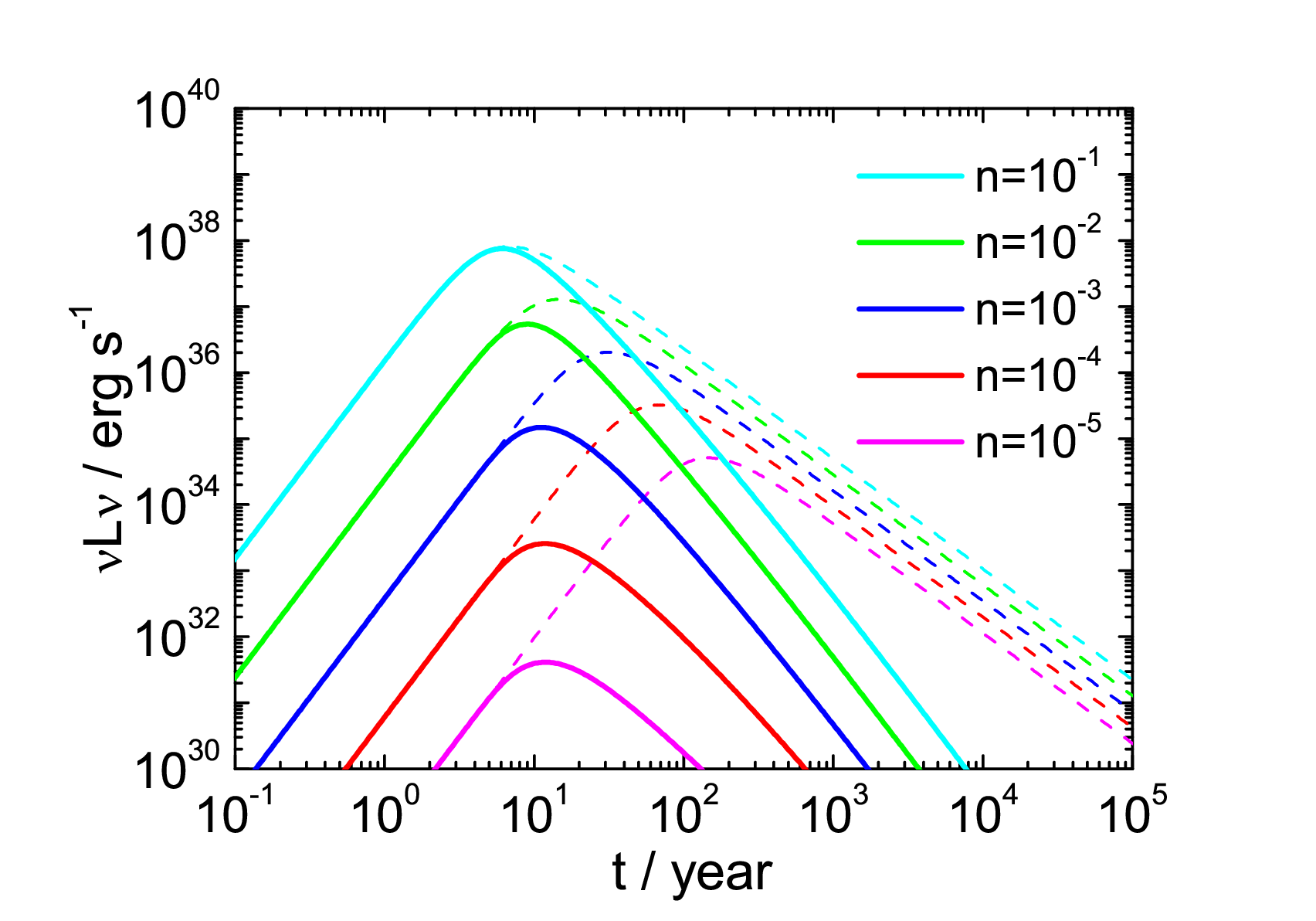}}\caption{The
radio (6GHz) lightcurves of complete-sweeping (dashed lines) and
incomplete-sweeping (solid lines) scenarios. The basic parameters
are the same with Fig.\ref{fig1} but with varied $n$.}\label{fig2}
\end{figure}

\begin{figure}[t]
\centering\resizebox{1.0\hsize}{!}{\includegraphics{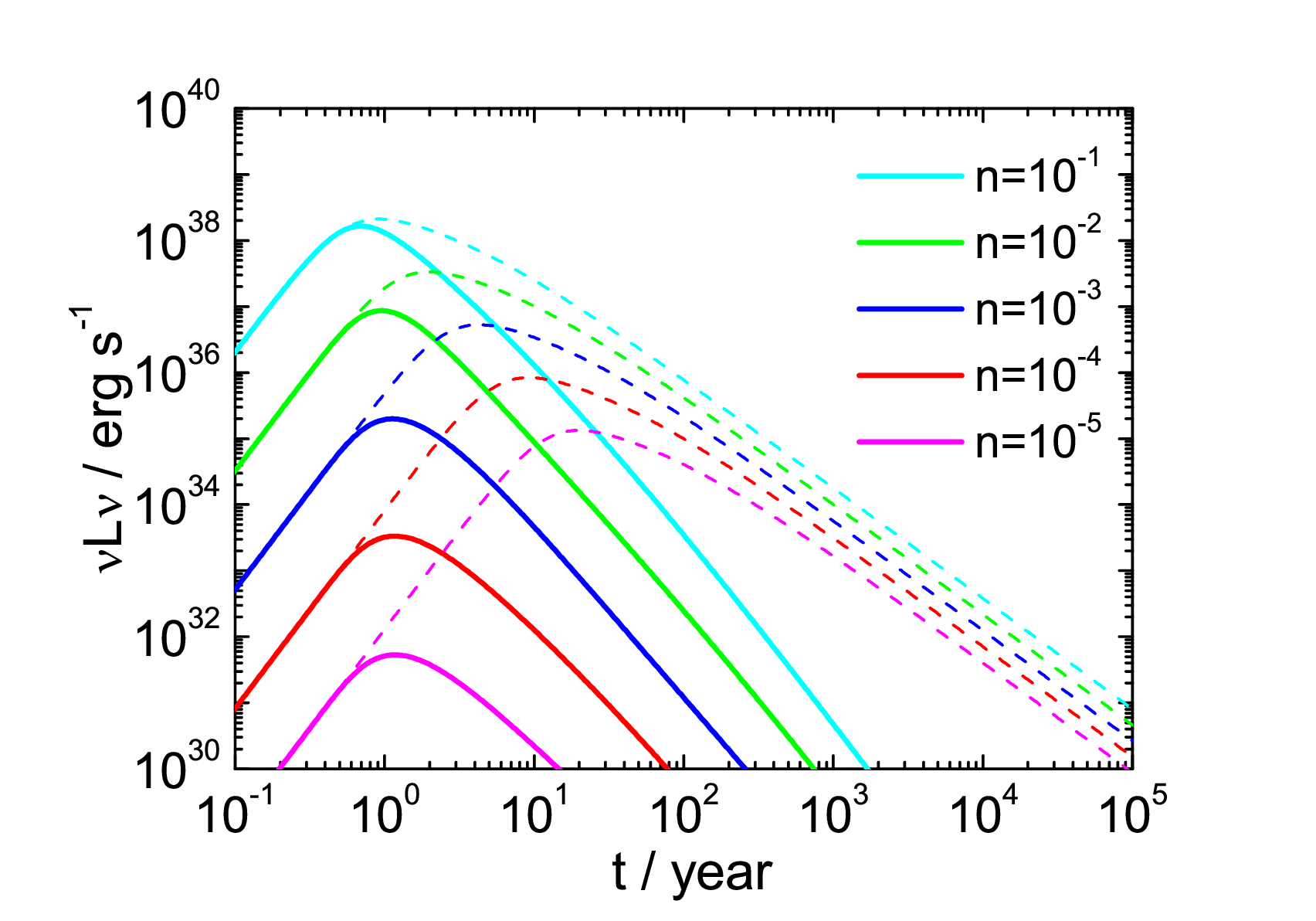}}\caption{The
same as Fig.\ref{fig2}, but with $M_{\rm ej}=0.001 M_{\odot}$ and
$\beta=0.6$.}\label{fig3}
\end{figure}

\begin{figure*}[t]
\centering\resizebox{1.0\hsize}{!}{\includegraphics{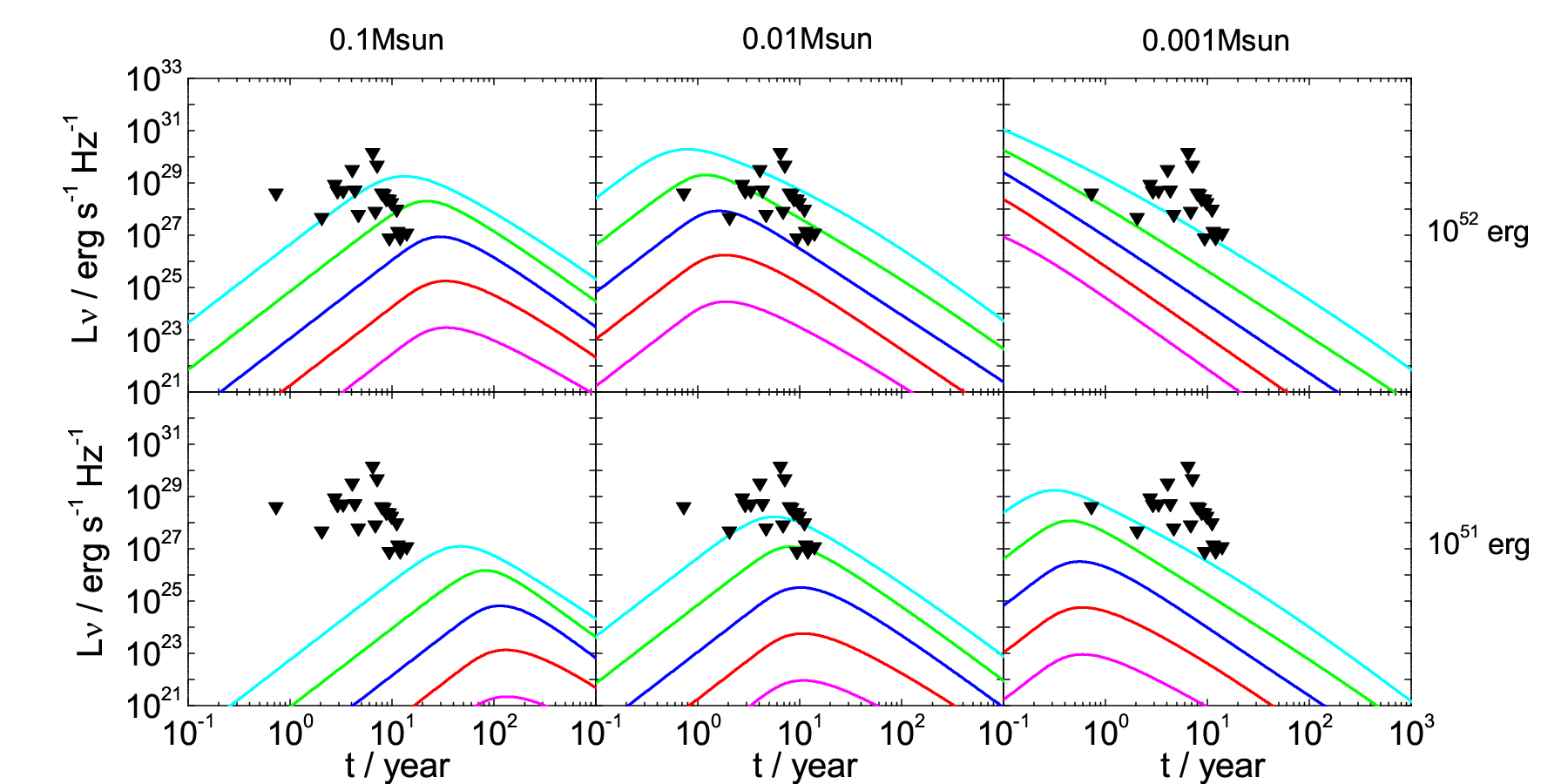}}\caption{The
radio lightcurves with different ejecta mass and explosion energy
and the comparison with observation. The upper limit data is from
Ricci et al. (2021). }\label{fig4}
\end{figure*}

\begin{figure*}[t]
\centering\resizebox{1.0\hsize}{!}{\includegraphics{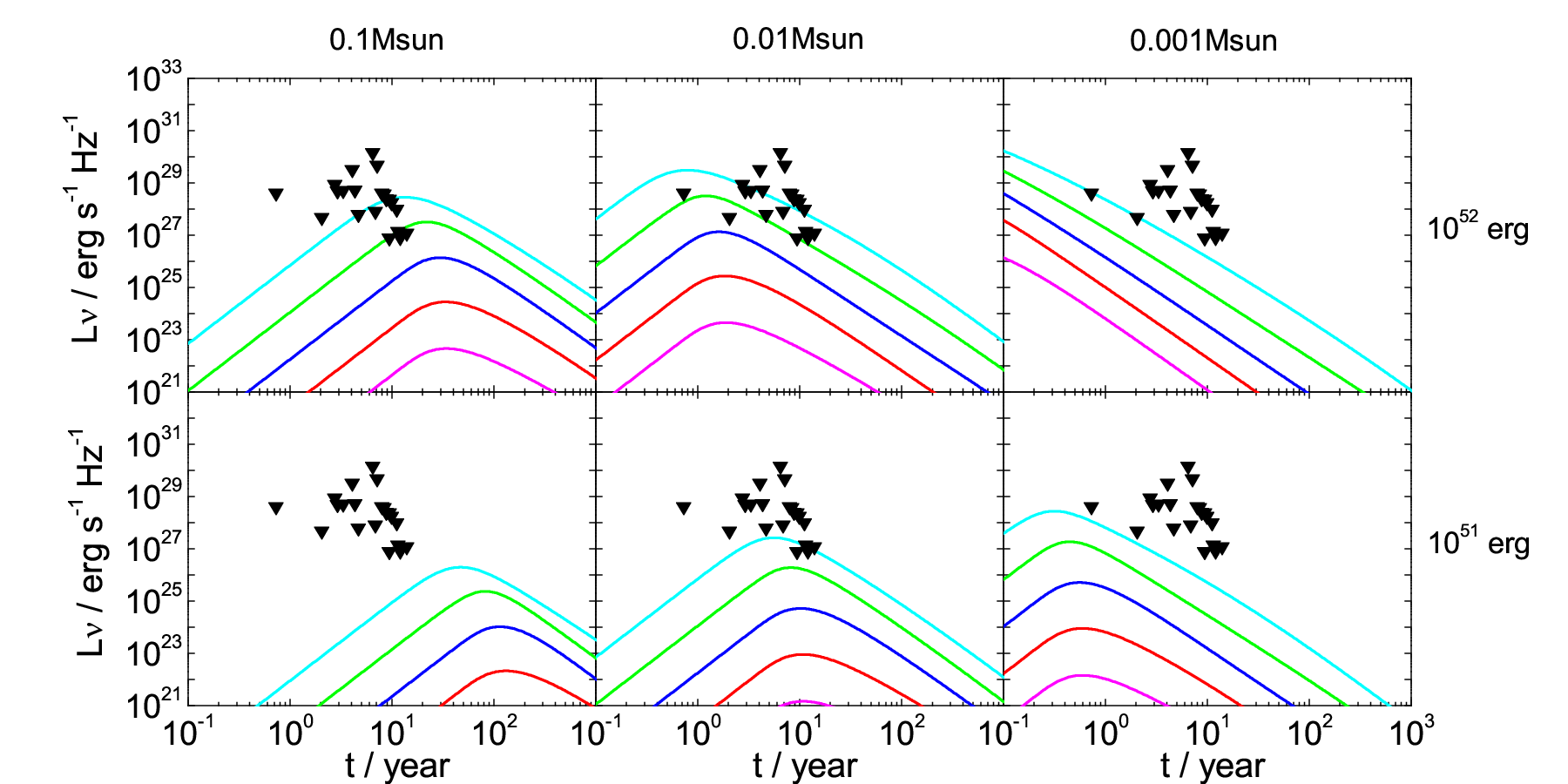}}\caption{The
same as Fig.3 but with $\epsilon_{\rm B}=0.01$.}\label{fig5}
\end{figure*}

\section{The emission}
At late time, the merger ejecta is too thin to sweep up the ISM.
Meanwhile, a global steady external shock is unfavored. However, the
local shock may exist somewhere, since the ejecta has a density
profile or fluctuation. Therefore, the radiation calculation in
Nakar \& Piran (2011) and Metzger \& Bower (2014) is still adopted
here, but with a little adjustment. The shocked electrons lose their
energy by synchrotron radiation. The flux and the luminosity to the
distant observer are
\begin{eqnarray}
F_{\nu} = {(1+z)N_{e} \over 4\pi D_{\rm L}^{2}}{m_{e}c^2\sigma_{\rm
T}\Gamma \beta_{\rm m}^2 B'\over 3e}{\left(\nu_{\rm obs}\over
\nu_{\rm m}\right)}^{-(p-1)/2}\label{Flux}
\end{eqnarray}
and
\begin{eqnarray}
L_{\nu} = F_{\nu}\, 4\pi D_{\rm L}^{2},
\end{eqnarray}
 where the typical synchrotron frequency of electrons is
$\nu_{\rm m}=3\Gamma \gamma_{\rm m}^2\beta_{\rm m}^2 e B'/4\pi
m_{e}c$, the minimum electron Lorentz factor is $\gamma_{\rm m}=
({(p-2)m_{\rm p}/ (p-1)m_{e}})\epsilon_{e}(\Gamma-1)+1$ (Liu et al.
2020), and $\nu_{\rm obs}>\nu_{\rm m},\nu_{\rm a}$ is adopted in
most realistic scenarios, following Nakar \& Piran (2011). The
parameters $p=2.2$ and $\epsilon_{e}=0.1$ are adopted in this paper.
The local magnetic field strength in the shock frame could be
estimated by
\begin{eqnarray}
B'= \sqrt{8\pi \tilde{\epsilon}_{\rm
B}(4\Gamma+3)(\Gamma-1)\tilde{{n}}_{\rm }m_{\rm p}c^2},
\end{eqnarray}
where the effective medium density $\tilde{n}=(1-e^{-\tau})n$.
Considering that the shock can not keep steady once $\tau\leqslant
1$, the shock-induced magnetic energy would decay following the
decrease of $\tau$. Here, a similar expression,
$\tilde{\epsilon}_{\rm B}=(1-e^{-\tau}){\epsilon}_{\rm B}$, is
adopted to characterize the magnetic energy fraction.

\begin{figure}
\centering\resizebox{1.0\hsize}{!}{\includegraphics{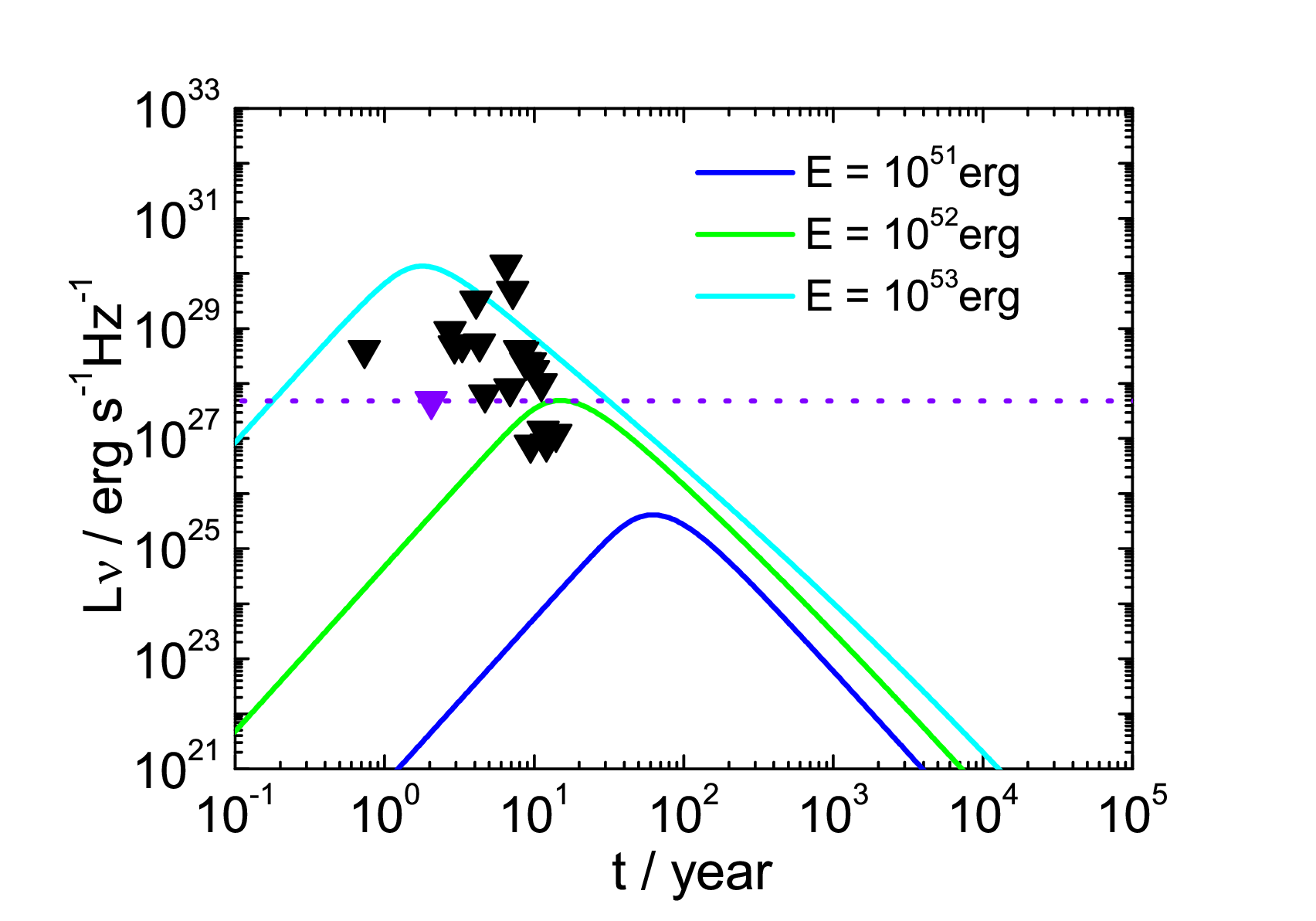}}\caption{Constraints
on the kinetic energy of the merger ejecta of GW170817. The ejecta
mass and medium density are fixed as $m_{\rm ej}=0.06\,M_{\odot}$
and $n=2\times10^{-3}$. The limit from GW170817 is marked in violet.
The limits from other GRB samples are shown in black for
comparison.}\label{fig6}
\end{figure}

\begin{figure}[t]
\centering\resizebox{1.0\hsize}{!}{\includegraphics{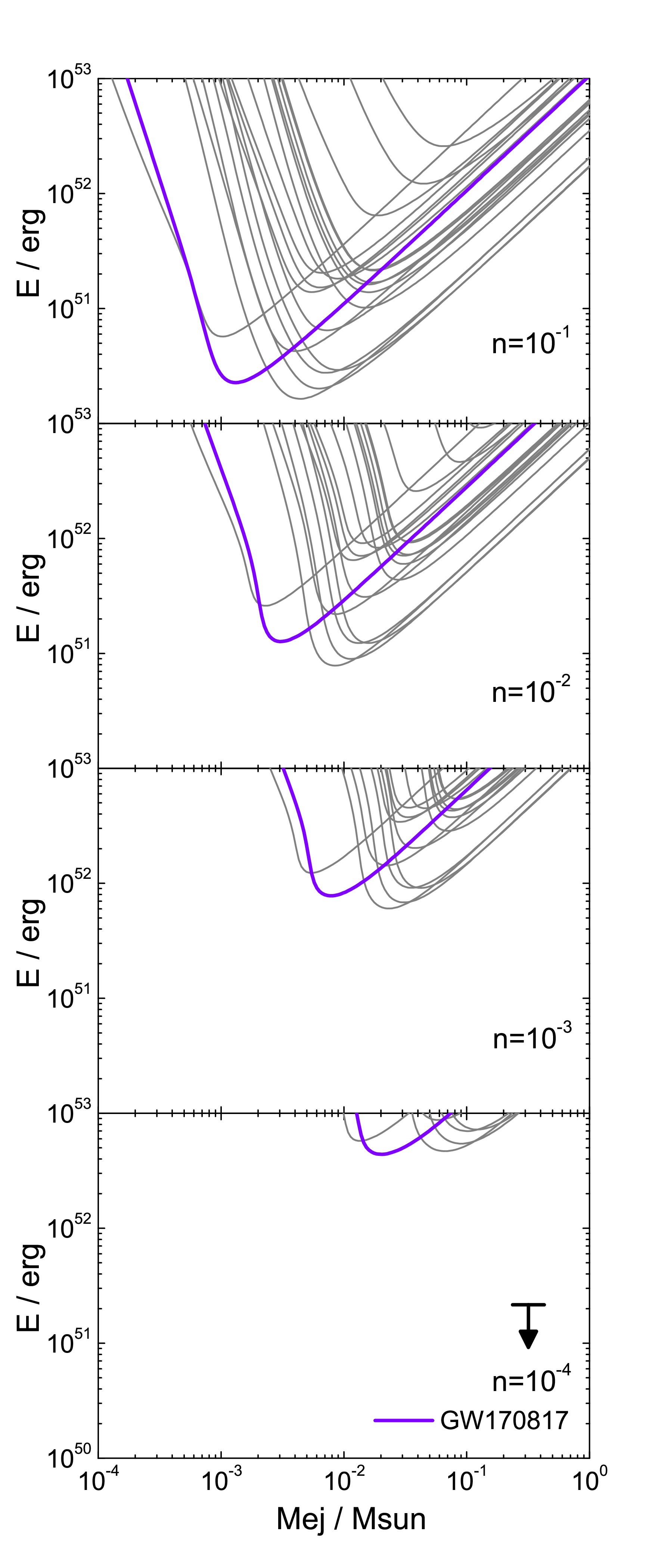}}
\caption{Upper limit lines of the constraints on the ejecta mass and
kinetic energy from 21 SGRB samples. The violet lines are from the
limit of the GW170817 event. }\label{fig7}
\end{figure}

As shown in Fig.\ref{fig1}, using typical ejecta parameters, we plot
the dynamics of the ejecta in both the complete-sweeping scenario
and the incomplete-sweeping scenario. The outcomes show much
difference at late time. In the complete-sweeping scenario (dashed
lines), the sweeping rate increases endlessly, even though the
ejecta has been sufficiently decelerated. But in the
incomplete-sweeping scenario (solid lines), the ejecta reaches
$\tau=1$ before deceleration. After that, when $\tau<1$, the ejecta
becomes thin and the sweeping rate begins to drop. On the one hand,
the incomplete sweeping leads to a lower sweeping rate; on the other
hand, the ejecta could keep its velocity for a much longer
timescale, as mentioned above. As a natural consequence, the energy
converted into radiation is decreased. Figure 2 shows the light
curves for both scenarios with different ISM densities. It is
independent of the medium density when the transition between $l< R$
and $l> R$ happens. The light curves begin to drop at the
transition. Therefore, the peak luminosity in the
incomplete-sweeping scenario should be at the transition time,
$t_{\rm peak}\simeq t_{\tau}$ ($\tau=1$), once the precondition
$n<n_{\rm c}$ is satisfied. We can see that for a higher medium
density, the difference of the peak luminosity in the
complete-sweeping and incomplete-sweeping scenarios is not clear.
But for a lower density, the peak luminosity in the
incomplete-sweeping scenario is much lower than that in the
complete-sweeping scenario. For a better understanding, we plot
Figure 3, using various ejecta parameters for comparison. In simple
terms, the lesser the ejecta mass and the lower the ISM density,
much more different peak luminosities will show in these two
scenarios. This is actually determined by Equation (10). It is more
sensitive to the parameter $n$ than $M_{\rm ej}$. However, from
Equation (9), the ejecta mass and velocity jointly determine the
transition timescale, which takes responsibility for the peak
luminosity.

The ejecta cannot sweep up enough medium after the phase transition,
and the condition $l\gg R$ for a steady shock is broken. Therefore,
the late-time radio emission could be much lower than the estimates
in previous studies. Strictly speaking, Equation (14) is
inapplicable when the shock is broken. For simplicity, we still
calculate the light curves using this equation, in case of what we
are concerned about is only the peaks of the light curve.
Furthermore, we plot the radio light curves using different initial
parameters and then take a comparison with the observation. As shown
in Figure 4, for an energetic explosion (e.g., $E_{\rm k}=10^{52}\rm
erg$) with a relatively higher ISM density, the light curves may
still exceed the upper limit from observations. This is consistent
with previous works (Metzger \& Bower 2014; Fong et al. 2016; Liu et
al. 2020). However, under a relatively lower ISM density, our
calculation shows that the light curves are obviously below the
upper limit. If the explosion is not powerful enough (e.g., $E_{\rm
k}=10^{51}\rm erg$), our results are in accord with the observation:
the shock emission is weak and the radio signals can hardly be
detected. Furthermore, our calculation pays less attention to the
specific radiation mechanism. According to short-GRB afterglows, a
relatively small value of $\epsilon_{\rm B}$ may be favored. It
would make the radio signal more undetectable, as shown in Figure 5
(see the detailed discussion in Liu et al. 2020).

\section{the constraints}
As the first detection of GW signals from double NS mergers, the
GW170817 event gives a chance to investigate the late-time radio
emission. First, from the multiband afterglow of GRB170817A, the ISM
density is limited as $n=2.5^{+4.1}_{-1.9}\times 10^{-3}\,\rm
cm^{-3}$ (Hajela et al. 2019). Second, if the associated kilonova
(AT2017gfo) is powered by the radioactive decay of r-process
elements, the total mass of the merger ejecta is estimated to be
roughly $0.06\,M_{\odot}$ (Smartt et al. 2017; Tanaka et al. 2017;
Tanvir et al. 2017; Villar et al. 2017). Adopting these two
parameters, Figure 6 is plotted to limit the kinetic energy of the
merger ejecta. It clearly shows that the late-time radio signal can
hardly be detected with either a total energy of $10^{52}\rm erg$
nor $10^{51}\rm erg$. In another case, the required ejecta mass
could be much less than $0.06\,M_{\odot}$ if AT2017gfo is powered by
an NS or magnetar (Ai et al. 2018; Li et al. 2018; Yu et al. 2018).
Therefore, we plot more general constraints on the ejecta mass and
kinetic energy in Figure 7. For completeness, all 21 samples from
Ricci et al. (2021) are included. The observational limits are
converted into upper-limit lines in this plot. One can see that it
provides various constraints on the ejecta mass and kinetic energy,
depending on the medium density. Each of the lines clearly shows a
decline and rise in the energy¨Cmass plot. The rise on the right
side of the lines corresponds to the complete-sweeping scenario and
is consistent with the results in Ricci et al. (2021). For the
decline on the left, it corresponds to the incomplete-sweeping
scenario. The watershed of the decline and rise is actually
determined by Equation (9), where the deceleration timescale and
transition timescale get a balance. From Figure 7, an energetic
explosion (e.g., $E_{\rm k}=10^{52}\rm erg$) is definitely allowed
in a broad parameter space, especially with low medium densities.

\section{conclusion and discussion}

Observing long-lasting radio emissions from merger events helps us
to understand the evolution of the merger ejecta. The radio emission
should be much brighter if the post-merger remnant is an NS instead
of a BH. However, based on the reasonable neutral ISM assumption,
our results show that the merger ejecta would be too rarefied to
sweep up the ISM several years after the explosion. Instead of
sweeping, the merger ejecta would eventually diffuse in the ISM,
losing its energy over a much longer timescale through atomic
collisions or scattering. As a result, the radio signal is too weak
to be detected under typical conditions. Even with an NS remnant, it
is still hard to catch the radio signal if the density of the ISM is
not high enough. Therefore, it is not easy to rule out an NS (or
magnetar) remnant with the current observations.

It is usually considered that the merger remnant is rotating near
the Kepler period because of its significant orbital angular
momentum. With some possible mechanisms, such as
magnetic-field-induced ellipticities (Bonazzola \& Gourgoulhon 1996;
Palomba 2001; Cutler 2002) and various instability modes (Lai \&
Shapiro 1995; Andersson 1998; Lindblom et al. 1998; Corsi \&
M\'{e}sz\'{a}ros 2009; Dall'osso et al. 2015), the GW radiation
becomes significant for such a rapidly rotating NS. Based on short
GRBs, Gao et al. (2016) demonstrate that the rotation energy of the
merger remnant is probably lost in the GW channel. L\"{u} et al.
(2020) find a short-GRB sample as evidence for the GW-dominated
afterglow emission. Therefore, it is most likely that only a
fraction of the energy could be reflected in the X-ray plateau and
kilonova emission, which is only on the order of $10^{51}\rm erg$
(Rowlinson et al. 2013; L\"{u} et al. 2015; Gao et al. 2016; Yu et
al. 2018; Li et al. 2018). With such a total energy, the
long-lasting radio signal is hard to detect. Anyhow, the first
detection of a post-merger GW signal is still expected, once the
advanced LIGO and Virgo reach their designed sensitivity (Abbott et
al. 2017c, 2019; Caudill et al. 2021). At that time, the GW
observation could help us identify the merger remnant.

\acknowledgements  This work is supported by the National Natural
Science Foundation of China (grant No. 12021003, 11833003,
12303044), National Key R\&D Program of China (2021YFA0718500), and
the National SKA Program of China (2020SKA0120300, 2022SKA0130100).


\begin{thebibliography}{99}

\bibitem[Abbott et al.(2017)]{2017PhRvL.119p1101A} Abbott, B.~P., Abbott, R., Abbott, T.~D., et al.\ 2017a, \prl, 119, 161101
\bibitem[Abbott et al.(2017)]{2017ApJ...851L..16A} Abbott, B.~P., Abbott, R., Abbott, T.~D., et al.\ 2017b, \apjl, 851, L16
\bibitem[Abbott et al.(2017)]{2017ApJ...848L..13A} Abbott, B.~P., Abbott, R., Abbott, T.~D., et al.\ 2017c, \apjl, 848, L13
\bibitem[Abbott et al.(2019)]{2019ApJ...875..160A} Abbott, B.~P., Abbott, R., Abbott, T.~D., et al.\ 2019, \apj, 875, 160
\bibitem[Ag{\"u}{\'\i} Fern{\'a}ndez et al.(2023)]{2023MNRAS.520..613A} Ag{\"u}{\'\i} Fern{\'a}ndez, J.~F., Th{\"o}ne, C.~C., Kann, D.~A., et al.\ 2023, \mnras, 520, 613
\bibitem[Ai et al.(2018)]{2018ApJ...860...57A} Ai, S., Gao, H., Dai, Z.-G., et al.\ 2018, \apj, 860, 57
\bibitem[Ai et al.(2020)]{2020ApJ...893..146A} Ai, S., Gao, H., \& Zhang, B.\ 2020, \apj, 893, 146
\bibitem[Ai et al.(2022)]{2022MNRAS.516.2614A} Ai, S., Zhang, B., \& Zhu, Z.\ 2022, \mnras, 516, 2614
\bibitem[Andersson(1998)]{1998ApJ...502..708A} Andersson, N.\ 1998, \apj, 502, 708
\bibitem[Bonazzola \& Gourgoulhon(1996)]{1996A&A...312..675B} Bonazzola, S. \& Gourgoulhon, E.\ 1996, \aap, 312, 675
\bibitem[Bruni et al.(2021)]{2021MNRAS.505L..41B} Bruni, G., O'Connor, B., Matsumoto, T., et al.\ 2021, \mnras, 505, L41
\bibitem[Caudill et al.(2021)]{2021MPLA...3630022C} Caudill, S., Kandhasamy, S., Lazzaro, C., et al.\ 2021, Modern Physics Letters A, 36, 2130022-458
\bibitem[Corsi \& M{\'e}sz{\'a}ros(2009)]{2009ApJ...702.1171C} Corsi, A. \& M{\'e}sz{\'a}ros, P.\ 2009, \apj, 702, 1171
\bibitem[Cowperthwaite et al.(2017)]{2017ApJ...848L..17C} Cowperthwaite, P.~S., Berger, E., Villar, V.~A., et al.\ 2017, \apjl, 848, L17
\bibitem[Cutler(2002)]{2002PhRvD..66h4025C} Cutler, C.\ 2002, \prd, 66, 084025
\bibitem[Dai \& Lu(1998)]{1998A&A...333L..87D} Dai, Z.~G. \& Lu, T.\ 1998a, \aap, 333, L87
\bibitem[Dai \& Lu(1998)]{1998PhRvL..81.4301D} Dai, Z.~G. \& Lu, T.\ 1998b, \prl, 81, 4301
\bibitem[Dall'Osso et al.(2015)]{2015ApJ...798...25D} Dall'Osso, S., Giacomazzo, B., Perna, R., et al.\ 2015, \apj, 798, 25
\bibitem[Falcone et al.(2007)]{2007ApJ...671.1921F} Falcone, A.~D., Morris, D., Racusin, J., et al.\ 2007, \apj, 671, 1921
\bibitem[Fong et al.(2016)]{2016ApJ...831..141F} Fong, W., Metzger, B.~D., Berger, E., et al.\ 2016, \apj, 831, 141
\bibitem[Gao et al.(2013)]{2013ApJ...771...86G} Gao, H., Ding, X., Wu, X.-F., et al.\ 2013, \apj, 771, 86
\bibitem[Gao et al.(2015)]{2015ApJ...807..163G} Gao, H., Ding, X., Wu, X.-F., et al.\ 2015, \apj, 807, 163
\bibitem[Gao et al.(2016)]{2016PhRvD..93d4065G} Gao, H., Zhang, B., \& L{\"u}, H.-J.\ 2016, \prd, 93, 044065
\bibitem[Gao et al.(2017)]{2017ApJ...837...50G} Gao, H., Zhang, B., L{\"u}, H.-J., et al.\ 2017, \apj, 837, 50
\bibitem[Hajela et al.(2019)]{2019ApJ...886L..17H} Hajela, A., Margutti, R., Alexander, K.~D., et al.\ 2019, \apjl, 886, L17
\bibitem[Heintz et al.(2018)]{2018MNRAS.479.3456H} Heintz, K.~E., Watson, D., Jakobsson, P., et al.\ 2018, \mnras, 479, 3456
\bibitem[Horesh et al.(2016)]{2016ApJ...819L..22H} Horesh, A., Hotokezaka, K., Piran, T., et al.\ 2016, \apjl, 819, L22
\bibitem[Hotokezaka et al.(2018)]{2018IJMPD..2742005H} Hotokezaka, K., Beniamini, P., \& Piran, T.\ 2018, International Journal of Modern Physics D, 27, 1842005
\bibitem[Klose et al.(2019)]{2019ApJ...887..206K} Klose, S., Nicuesa Guelbenzu, A.~M., Micha{\l}owski, M.~J., et al.\ 2019, \apj, 887, 206
\bibitem[Lai \& Shapiro(1995)]{1995ApJ...442..259L} Lai, D. \& Shapiro, S.~L.\ 1995, \apj, 442, 259
\bibitem[Ledoux et al.(1998)]{1998A&A...337...51L} Ledoux, C., Petitjean, P., Bergeron, J., et al.\ 1998, \aap, 337, 51
\bibitem[Li et al.(2018)]{2018ApJ...861L..12L} Li, S.-Z., Liu, L.-D., Yu, Y.-W., et al.\ 2018, \apjl, 861, L12
\bibitem[Lindblom et al.(1998)]{1998PhRvL..80.4843L} Lindblom, L., Owen, B.~J., \& Morsink, S.~M.\ 1998, \prl, 80, 4843
\bibitem[Liu et al.(2020)]{2020ApJ...890..102L} Liu, L.-D., Gao, H., \& Zhang, B.\ 2020, \apj, 890, 102
\bibitem[L{\"u} et al.(2015)]{2015ApJ...805...89L} L{\"u}, H.-J., Zhang, B., Lei, W.-H., et al.\ 2015, \apj, 805, 89
\bibitem[L{\"u} et al.(2020)]{2020ApJ...898L...6L} L{\"u}, H.-J., Yuan, Y., Lan, L., et al.\ 2020, \apjl, 898, L6
\bibitem[Margalit \& Metzger(2017)]{2017ApJ...850L..19M} Margalit, B. \& Metzger, B.~D.\ 2017, \apjl, 850, L19
\bibitem[Margalit \& Piran(2020)]{2020MNRAS.495.4981M} Margalit, B. \& Piran, T.\ 2020, \mnras, 495, 4981
\bibitem[Metzger et al.(2017)]{2017ApJ...841...14M} Metzger, B.~D., Berger, E., \& Margalit, B.\ 2017, \apj, 841, 14
\bibitem[Metzger et al.(2011)]{2011MNRAS.413.2031M} Metzger, B.~D., Giannios, D., Thompson, T.~A., et al.\ 2011, \mnras, 413, 2031
\bibitem[Metzger \& Piro(2014)]{2014MNRAS.439.3916M} Metzger, B.~D. \& Piro, A.~L.\ 2014, \mnras, 439, 3916
\bibitem[Metzger \& Bower(2014)]{2014MNRAS.437.1821M} Metzger, B.~D. \& Bower, G.~C.\ 2014, \mnras, 437, 1821
\bibitem[Nakar \& Piran(2011)]{2011Natur.478...82N} Nakar, E. \& Piran, T.\ 2011, \nat, 478, 82
\bibitem[Nicholl et al.(2017)]{2017ApJ...848L..18N} Nicholl, M., Berger, E., Kasen, D., et al.\ 2017, \apjl, 848, L18
\bibitem[Palomba(2001)]{2001A&A...367..525P} Palomba, C.\ 2001, \aap, 367, 525
\bibitem[Pian et al.(2017)]{2017Natur.551...67P} Pian, E., D'Avanzo, P., Benetti, S., et al.\ 2017, \nat, 551, 67
\bibitem[Radice et al.(2018)]{2018ApJ...852L..29R} Radice, D., Perego, A., Zappa, F., et al.\ 2018, \apjl, 852, L29
\bibitem[Rezzolla et al.(2018)]{2018ApJ...852L..25R} Rezzolla, L., Most, E.~R., \& Weih, L.~R.\ 2018, \apjl, 852, L25
\bibitem[Ricci et al.(2021)]{2021MNRAS.500.1708R} Ricci, R., Troja, E., Bruni, G., et al.\ 2021, \mnras, 500, 1708
\bibitem[Rosswog et al.(2014)]{2014MNRAS.439..744R} Rosswog, S., Korobkin, O., Arcones, A., et al.\ 2014, \mnras, 439, 744
\bibitem[Rowlinson et al.(2013)]{2013MNRAS.430.1061R} Rowlinson, A., O'Brien, P.~T., Metzger, B.~D., et al.\ 2013, \mnras, 430, 1061
\bibitem[Sarin \& Lasky(2021)]{2021GReGr..53...59S} Sarin, N. \& Lasky, P.~D.\ 2021, General Relativity and Gravitation, 53, 59
\bibitem[Sarin et al.(2022)]{2022MNRAS.516.4949S} Sarin, N., Omand, C.~M.~B., Margalit, B., et al.\ 2022, \mnras, 516, 4949
\bibitem[Schroeder et al.(2020)]{2020ApJ...902...82S} Schroeder, G., Margalit, B., Fong, W.-. fai ., et al.\ 2020, \apj, 902, 82
\bibitem[Shibata et al.(2019)]{2019PhRvD.100b3015S} Shibata, M., Zhou, E., Kiuchi, K., et al.\ 2019, \prd, 100, 023015
\bibitem[Shibata et al.(2017)]{2017PhRvD..96l3012S} Shibata, M., Fujibayashi, S., Hotokezaka, K., et al.\ 2017, \prd, 96, 123012
\bibitem[Smartt et al.(2017)]{2017Natur.551...75S} Smartt, S.~J., Chen, T.-W., Jerkstrand, A., et al.\ 2017, \nat, 551, 75
\bibitem[Tanaka et al.(2017)]{2017PASJ...69..102T} Tanaka, M., Utsumi, Y., Mazzali, P.~A., et al.\ 2017, \pasj, 69, 102
\bibitem[Tanvir et al.(2017)]{2017ApJ...848L..27T} Tanvir, N.~R., Levan, A.~J., Gonz{\'a}lez-Fern{\'a}ndez, C., et al.\ 2017, \apjl, 848, L27
\bibitem[Tanvir et al.(2019)]{2019MNRAS.483.5380T} Tanvir, N.~R., Fynbo, J.~P.~U., de Ugarte Postigo, A., et al.\ 2019, \mnras, 483, 5380
\bibitem[Troja et al.(2017)]{2017Natur.551...71T} Troja, E., Piro, L., van Eerten, H., et al.\ 2017, \nat, 551, 71
\bibitem[Vaughan et al.(2006)]{2006ApJ...638..920V} Vaughan, S., Goad, M.~R., Beardmore, A.~P., et al.\ 2006, \apj, 638, 920
\bibitem[Villar et al.(2017)]{2017ApJ...851L..21V} Villar, V.~A., Guillochon, J., Berger, E., et al.\ 2017, \apjl, 851, L21
\bibitem[Wolfe \& Prochaska(2000)]{2000ApJ...545..591W} Wolfe, A.~M. \& Prochaska, J.~X.\ 2000, \apj, 545, 591
\bibitem[Yu et al.(2015)]{2015ApJ...806L...6Y} Yu, Y.-W., Li, S.-Z., \& Dai, Z.-G.\ 2015, \apjl, 806, L6
\bibitem[Yu et al.(2018)]{2018ApJ...861..114Y} Yu, Y.-W., Liu, L.-D., \& Dai, Z.-G.\ 2018, \apj, 861, 114
\bibitem[Yu et al.(2013)]{2013ApJ...776L..40Y} Yu, Y.-W., Zhang, B., \& Gao, H.\ 2013, \apjl, 776, L40
\bibitem[Zhang \& M{\'e}sz{\'a}ros(2001)]{2001ApJ...552L..35Z} Zhang, B. \& M{\'e}sz{\'a}ros, P.\ 2001, \apjl, 552, L35














\end{thebibliography}
\end{document}